# MAC protocol classification in the ISM band using machine learning methods

Hanieh Rashidpour, Hossein Bahramgiri

[hanierp@mut.ac.ir](hanierp@mut.ac.ir)

[Bahramgiri@mut.ac.ir](Bahramgiri@mut.ac.ir)

## Abstract

With the emergence of new technologies and a growing number of wireless networks, we face the problem of radio spectrum shortages. As a result, identifying the wireless channel spectrum with the goal of exploiting the channel's idle state while also boosting network security is pivotal issue. Detecting and classifying protocols in the MAC sublayer enables Cognitive Radio (CR) users to improve spectrum utilization and minimize potential interference. In this paper, we classify the Wi-Fi and Bluetooth protocols, which are the most widely used MAC sublayer protocols in the ISM radio band. With the advent of various wireless technologies, especially in the 2.4 GHz frequency band, the ISM frequency spectrum has become a crowded and high-traffic band, which faces a lack of spectrum resources and user interference. Therefore, identifying and classifying protocols is an effective and useful method. Leveraging machine learning (ML) and deep learning (DL) techniques, known for their advanced classification capabilities, we apply Support Vector Machine (SVM) and K-Nearest Neighbors (KNN) algorithms, which are machine learning algorithms, to classify protocols into three classes: Wi-Fi, Wi-Fi Beacon, and Bluetooth. To capture the signals, we use the USRP N210 Software Defined Radio (SDR) device and sample the real data in the indoor environment in different conditions of the presence and absence of transmitters and receivers for these two protocols. By assembling this dataset and studying the time and frequency features of the protocols, we extract the frame width and the silence gap between the two frames as time features and the PAPR of each frame as a power feature. By comparing the output of the protocols classification in different conditions and also adding Gaussian noise, it was found that the samples in the nonlinear SVM method with RBF and KNN functions have the best performance, with 97.83% and 98.12% classification accuracy, respectively.

## I. Introduction

With the rapid expansion of new wireless communication systems, including wireless personal area networks (WPANs), wireless local area networks (WLANs), and wireless metropolitan area networks (WMANs), the demand for radio spectrum is increasing day by day, so spectrum awareness and allocation become a more important subject. However, measurements in this area indicate that the wide range of spectrum allocated to users (primary networks) is underutilized. As a result, wireless channel spectrum identification with the goal of exploiting the channel's idle state and its optimal management, as well as network security, is a critical issue. Users can adjust their transmission parameters in Cognitive Radio (CR) systems by sensing the current state of the external radio environment [1], thus improving spectrum utilization and effectively reducing the problem of spectrum resource scarcity.

Spectrum utilization can be improved if the network is aware of channel parameters such as empty capacity, interference, signal modulation, media access control protocols (MAC), power levels, transmission schemes, and so on [2]. In particular, CR users can identify present protocols in any transmission by identifying and classifying MAC sublayer characteristics. This issue is more critical in heterogeneous spectrums such as the ISM (Industrial, Scientific, and Medical) bands. The 2.4 GHz ISM band is unlicensed and free, and several important technologies like Wi-Fi, ZigBee, and Bluetooth share it. As a result, it has become a crowded and high-traffic band, facing user and network interferences. These technologies compete for resources and strive to coexist. On the other hand, in city monitoring systems performed by government institutions, it is necessary to identify and review active protocols in the electromagnetic spectrum. This application also amplifies the significance of wireless protocol identification and classification, especially in crowded bands.

Many researchers have utilized machine learning (ML) and deep learning (DL), the most advanced classification tools, to identify and classify active technologies in the radio spectrum. Many DL models have been developed for various tasks, including Convolutional Neural Networks (CNN). In [3], using the CNN method, the authors classify Wi-Fi, Bluetooth, and Microwave signals, which are ISM band signals, into eight different classes. Therefore, the wireless network can adjust its transmission patterns, such as power selection and channels, if it detects and identifies certain types of signals. Convolution neural network (CNN) and support vector machine (SVM) in the study of identification of MAC protocols [4],[2], signal detection in mobile cellular networks [5],[6], modulation classification [7],[8], and sensing the spectrum [9] have achieved extraordinary performance. In [10], the authors classify the traffic of homogeneous and heterogeneous networks 802.11b/g/n using the K Nearest Neighborhood (KNN) method, which is one of the ML approaches. The SVM, as one of the machine learning methods, has been widely used as a classification method for spectrum measurement, modulation classification, user identification, and power allocation in cognitive radio. In [4], Hu *et al*, study the identification of MAC protocols for applications in cognitive systems. Four protocols, TDMA, CSMA/CA, Slotted Aloha, and Pure Aloha are identified and classified using the SVM method. They try to show that, in addition to being effective in decreasing interference and spectrum shortage, spectrum sensing can also be used to communicate amongst heterogeneous cognitive radio networks. In [4], power features like mean and variance of received signal power and time features like channel idle state duration and channel busy state duration are extracted, and the performance of kernel functions in different conditions are compared. [11] uses and compares four linear classification methods to classify Bluetooth and Wi-Fi protocols. By using a novel and low-cost technique based on feature extraction, [12] successfully identifies Wi-Fi, Bluetooth, and ZigBee protocols with a success rate of above 90%.

In this paper, with the goal of solving a practical and developable problem, we consider two protocols: Wi-Fi and Bluetooth. We use the USRP N210 to obtain real signals of these protocols and sample them in the 2.4 GHz ISM band in an indoor environment. We apply an approach in the time domain to automatically extract the signal bursts or frames. Compiling this dataset while examining the time and power features of the protocols, we extract the frame width and the silence gap between the two frames as time features and the peak-to-average power ratio (PAPR) of each frame as a power feature. Specifically, in this paper, we use the Support Vector Method (SVM) and the K Nearest Neighborhood (KNN) to classify protocols into three classes: Wi-Fi, Wi-Fi Beacon, and Bluetooth. For a more thorough analysis, the prepared signals from the protocols are infused with Gaussian noise to investigate its effect on classification techniques' performance. Finally, the performance of methods is evaluated, and the optimum method in terms of classification accuracy is proposed.

The structure of this article will be as follows. In Section II, we explain briefly the two target protocols, Wi-Fi and Bluetooth, and also two machine learning methods, SVM and KNN. An experimental approach to collect data by USRP N210 and check the output signal will be explained in Section III. In Section IV, feature extraction is explained, and then protocol classification algorithms are presented. Also, the results of the experiments are discussed. Finally, in Section V, we summarize and conclude the paper and also provide suggestions for future research.

## II. Preliminaries

The SVM and the KNN [13], [14] are two basic machine learning methods that can classify datasets into several classes. In this article, we use these methods to identify and classify Wi-Fi and Bluetooth protocols. Therefore, we will first briefly explain the characteristics of these two protocols and these two machine-learning methods.

### A. Wi-Fi (Specifically 802.11n)

IEEE 802.11 is a set of MAC channel access control specifications and a PHY physical layer for wireless LAN (WLAN) communications [15]. Due to the availability and currency of the 802.11n standard, we have used this standard in our work [16],[17]. DSSS modulation was first used for the 802.11 standards, while OFDM was used in many later standards, including 802.11n [18]. Instead of using a fast-modulated broadband signal, OFDM uses several slowly modulated narrowband signals to facilitate channel tuning [16]. In fact, OFDM enables high-speed data transfer by splitting data into several layered parallel bit streams, which are modulated on several separate subcarriers. The MAC sublayer provides performance tools for data transfer between networks as well as error detection and correction at the physical layer. This sublayer is responsible for queuing up nodes on a shared channel, using different protocols, each with its own techniques, to prevent collisions. This sublayer has an optional approach to reduce collisions between nodes using RTS and CTS frames. Before sending the data frame, the source node sends the RTS frame as the first step to notify the destination node and other nodes that it wants to send a data frame. The destination sends a CTS frame back to the source node. The CTS provides collision control management for a period of time so that other nodes are notified during the transmission that the channel is busy, and they stop any transmission in the channel. The exchange of frames is shown in Figure 1.

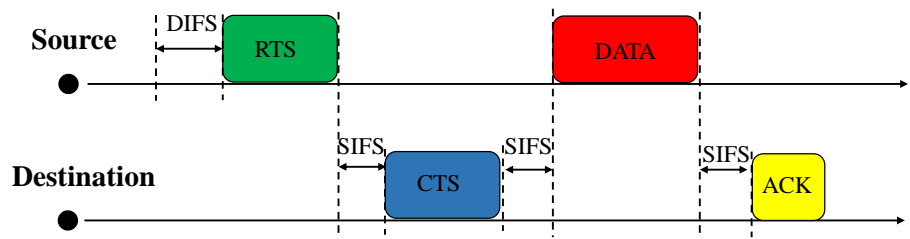

**Figure 1.** IEEE 802.11 channel access using RTS/CTS frames.

The destination sends an ACK frame to the source after receiving data from the source if the data is secure and no errors are detected. As shown in Figure 1, each frame is sent and received at a specified and determined time. Distributed coordination function Inter-Frame Space (DIFS) refers to the time a node must wait before sending its request frame. Short Inter-Frame Space (SIFS) is the time it takes to receive and respond to a frame in microseconds to process it with a single frame. Depending on the physical layer used, SIFS time includes RF receiver delay and MAC processing delay [19]. In each 802.11 standard, SIFS has a different time, which in 802.11n is equal to $10\mu s$.

### B. Wi-Fi Beacon

A beacon is sent from the Access Point (AP) according to the 802.11 standards, periodically to announce its presence and provide the Service Set Identifier (SSID) or network name to the devices that want to connect to the network to be informed of the status of the network. Table 1 shows the time period of each standard. As can be seen, the time size of each beacon frame varies according to the base rate.

**Table 1.** Frame time of Wi-Fi Beacon [20]

| Beacon Interval | Beacon Airtime | Basic Rates | Mode |
|---|---|---|---|
| 102.4 ms | 2.232 ms | 1, 2, 5.5, 6, 11, 12, 24 Mbps | b/g/n/ac |
| 102.4 ms | 1.896 ms | 1, 2, 5.5, 11 Mbps | b/g/n |
| 102.4 ms | 0.336 ms | 6, 12, 24 Mbps | Ac |
| 102.4 ms | 2.648 ms | 1, 2, 5.5, 6, 11, 12, 24 Mbps | b/g/n/ac |
| 102.4 ms | 2.184 ms | 1, 2, 5.5, 11 Mbps | b/g/n |
| 102.4 ms | 0.464 ms | 6, 12, 24 Mbps | Ac |

## C. Bluetooth (802.15 Standard)

Over time, the demand for access to mobile phones and wireless LAN networks led to the growth and development of Bluetooth technology [21]. The benefits and rapid proliferation of LANs suggest that establishing Personal Area Networks (PANs), that is, communications between devices in close proximity to the user, will have useful applications. Bluetooth network technology is designed with characteristics that are suited for PAN networks, such as short-range, low power consumption, small protocol stack, simultaneous audio and data transmission, and low cost. The main unit of a Bluetooth network is a piconet, which consists of one *master* device as a transmitter and one to seven *slave* devices as receivers. The Gaussian Frequency Shift-Keying Modulation (GFSK) modulation is used in this standard to maximize bandwidth utilization. By dividing the total bandwidth into 79 physical channels, each with a bandwidth of 1MHz, Bluetooth implements the frequency-hopping spread spectrum (FHSS) technique. This technique occurs by hopping in a sequence from one physical channel to another. The device designated as the master device of a piconet determines the FH sequence as a function of address, and the slave device follows the frequencies synchronized with the master and receives the message. The hop rate is 1600 times per second, so each physical channel is occupied for a period of $625\mu s$. The receiver responds to the frame on the same RF channel the transmitter uses. Depending on the data length, these forms can be considered by occupying 1, 3, or 5 time slots [21]. The length of each of these slots is a fixed duration in the Bluetooth standard, given in Table 2. In the Bluetooth standard, for example, ACK frames are sent between 126 and $366\mu s$.

**Table 2.** Bluetooth Standard Specification [21].

| Slot | Fixed Duration | Min Duration | Max Duration |
|---|---|---|---|
| Time Slot | $625\mu s$ | | |
| 1-Time Slot | | $126\ \mu s$ | $366\ \mu s$ |
| 3-Time Slot | | $1250\ \mu s$ | $1622\ \mu s$ |
| 5-Time Slot | | $2500\ \mu s$ | $2870\ \mu s$ |
| Null Packet | $126\ \mu s$ | | |

Like the 802.11 standard, the receiver sends an ACK frame to the transmitter in response to the successful receipt of the correct data if no error is detected. At this time, the transmitter makes sure that the receiver receives the correct data and sends the next data. In the next section, when sampling the Bluetooth protocol signal, we see the ACK frames that are sent about $200\mu s$ after the data frame is sent in the frequency domain.

## D. Support Vector Machines (SVM)

In 1919, Vapnik introduced support vector machines [22]. In its simplest model, linear SVM, it is a line that separates a set of separable samples into two classes with a maximum distance. We can separate the samples in a two-dimensional feature space with a line, and we do the same in a three-dimensional feature space using a plane. Therefore, this problem can generally be considered in the $n$-dimensional property space, which uses the $(n-1)$-dimensional hyperplane. In other words, a plane with the maximum distance from the classes is appropriate. This condition is also interpreted as having a maximum margin [13]. In general, for a hyperplane separator, it is defined,

$$\sum_i w_i x_i + b = 0, \tag{1}$$

or on the vector representation, it can be expressed as,

$$u = \mathbf{w}.\mathbf{x} + b, \tag{2}$$

Where **w** is the weighing vector perpendicular to the hyperplane, and b is a constant. In this view, u = 0 refers to the hyperplane itself, and the nearest points are on the page u = ± 1. Therefore, space is divided into two classes of samples with the properties of equations (4) and (5):

$$x_i.w + b \geq +1 - \xi_i \quad for \ y_i = 1 \tag{4}$$
$$x_i.w + b \geq -1 + \xi_i \quad for \ y_i = -1 \tag{5}$$

Because the samples are noisy and have certain errors, the variable "$\xi_i$" is defined as a slack. In this case, the optimization problem converts to finding w such that equation (7) is reduced:

$$\text{subject to } y_i(w.x_i + b) \geq 1 - \xi_i \ \ '\xi_i > 0 \min \frac{1}{2}\|w\|^2 + C \sum_i \xi_i \tag{7}$$

When the data points are not linearly separable in the data space, the SVM uses kernel functions to separate these data points. SVM maps the data to a transformed feature space and separates it linearly by the hyperplane using kernel functions so that every sample is replaced by a nonlinear kernel function $x \to \varphi(x)$, so we have:

$$k(x_i, x_j) \to \varphi(x_i).\varphi(x_j) \tag{8}$$

The above conditions indicate a linear separation but in higher dimensions. The following are the three most commonly used kernel functions [14]:

Linear Kernel:
$$k(x_i, x_j) = x_i^T x_j \tag{9}$$

Polynomial kernel:
$$k(x_i, x_j) = (x_i, x_j + c)^p \tag{10}$$

Gaussian radial basis kernel (RBF):
$$k(x_i, x_j) = \exp(\frac{-\|x_i - x_j\|^2}{c}) \tag{11}$$

## E. K Nearest Neighbor (KNN)

The nearest neighbor method is a sample-based method. This method does not use the data as a training dataset but rather saves it, and whenever a new test is presented, it examines its relationship to the previously saved dataset. Finally, it recognizes which class the new test belongs to. This method is used to represent the new test as a vector in the property space, and the Euclidean distance of the test vector is calculated with all the training vectors. Then, *K* samples, which are the nearest neighbors to the test point, are selected. After identifying the neighboring points, the class of the test point is determined by a majority vote. [13],[14]. Of course, several methods exist to classify a test vector, and the traditional nearest neighbor K algorithm determines a new test based on the number with the most votes. In this method, it is assumed that all samples are in a space of dimensions *n* at $R^n$, and each desirable sample **x** is represented by the following vector:$\mathbf{x} = <a_1(\mathbf{x}), a_2(\mathbf{x}), \ldots, a_n(\mathbf{x})>$ Where $a_r(\mathbf{x})$ represents the $r^{th}$ attribute of the sample **x**. The Euclidean distance between the two samples $\mathbf{x}_i$ and $\mathbf{x}_j$, defined by $d(\mathbf{x}_i, \mathbf{x}_j)$, is

$$d(\mathbf{x}_i, \mathbf{x}_j) = \sqrt{\sum_{r=1}^{n} \left(a_r(\mathbf{x}_i) - a_r(\mathbf{x}_j)\right)^2}. \tag{12}$$

After determining the *K* number of the nearest neighbor, we vote to find out which nearest neighbor's tag is the most, relate it to that new test, and classify it in that class. The following voting relation expresses this concept:

$$\hat{f}(x_q) \leftarrow arg\ max \sum_{i=1}^{k} \delta(v, f(x_i)) \tag{13}$$

## III. Experimentation

Today, software-defined radio (SDR) projects consist of two parts: software and hardware. Users must design both software and hardware to build the desired system. USRP family is one of the most common hardware used in various software-defined radio projects. Users can utilize the GNU Radio software's libraries to communicate with the USRP family of hardware and connect to other hardware. Here, we use the USRP N210 and GNU Radio software to sample the signals. The USRP N210 can stream up to 50 MS/s to and from host applications. This setup can sample 8-bit data with a rate of 50 MS/s or 16-bit data with a rate of 25MS/s. The SBX daughter board is placed inside the device for the desired operating frequency. Therefore, we consider the ISM band for taking Wi-Fi and Bluetooth signals in the frequency range of 2.400 to 2.480GHz.

### A. Sampling Wi-Fi Beacon signal

A beacon frame is a management frame in IEEE 802.11 WLANs. It contains all the information about the network so that other devices are notified of the network status. The access point (AP) periodically transmits beacon signals every 102.4ms. The AP shown in Figure 2 transmits only the 802.11n beacon signal on a specific channel. According to the AP's capability, we set the channel bandwidth to 20MHz. Thus, the sampling rate to receive the correct signal must be at least 20MSamples/sec.

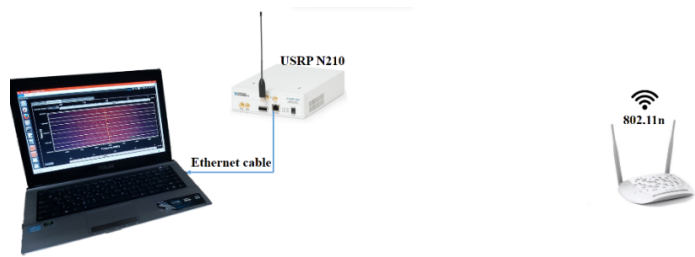

**Fig2.** The experiment setup.

We use the MATLAB software to see the details of the sampled signals. The width of the beacon frame, as indicated in Figure 3(a), is 2.184ms (the same as one of the options in Table 1). Figure 3(b) shows that a beacon signal is transmitted every 102.4ms. These two figures show limited parts of a Wi-Fi modem while it was in a beacon sending status in time and time-frequency domains.

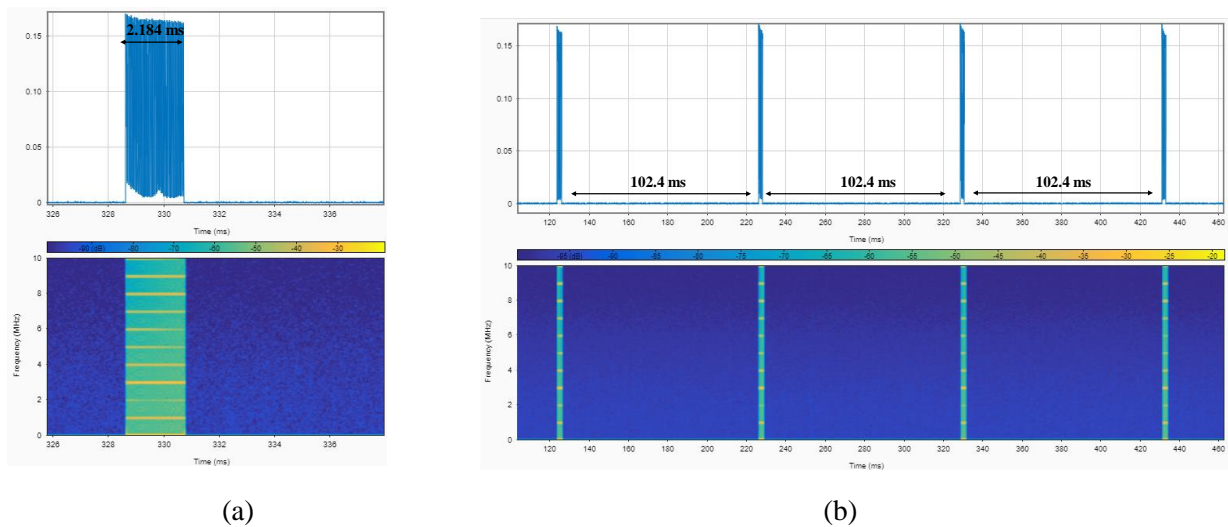

(a)  (b)

**Fig3.** (a) The beacon frame width. (b)The gap between the beacon frames.

### B. Sampling Wi-Fi signal

In this case, we consider the AP and a device that is connected to the AP and downloads large files from the Internet through the AP. Interestingly, we can see the steps of Figure 1in this signal, as shown in Figure 4. Figure 4(a) shows Wi-Fi signal during the download process in the time-frequency domain in GNU radio software, while Figure 4(b) shows a more limited part of the signal in the time domain and in MATLAB.

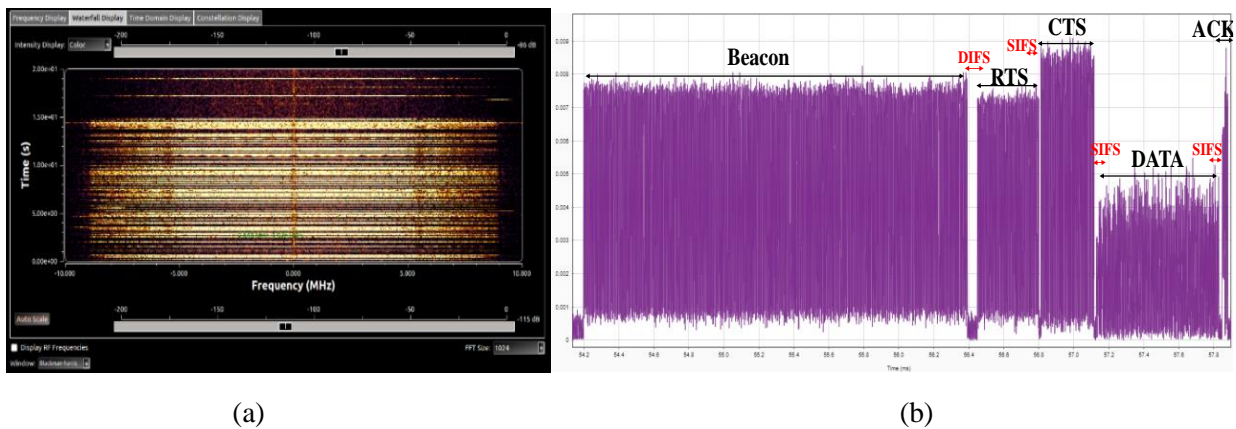

(a)  (b)

**Fig4.** (a) Wi-Fi Data signal in GNU Radio. (b) Wi-Fi Data signal in MATLAB in the time domain

In this part of the signal, the beacon frame precedes (accidentally) the RTS frame. The source transmits the RTS frame to the destination after 50 microseconds (DIFS), and after $10\mu s$ (SIFS), it responds to the source with a CTS frame. The AP then senses the CTS frame and, after $10\mu s$ (SIFS), sends the data frame. The destination sends the ACK frame after $10\mu s$ to notify the successful receiving of the data frame. Because the file size is large and the network has more data to transmit, these steps are repeated throughout the signal. Fig. 5 displays a larger part of the signal in the time and time-frequency domains. The RTS, CTS, and ACK frames carry side information about the data frame. Because of this, in the frequency-time domain, these frames do not use all of their subcarriers (compared to data frames).

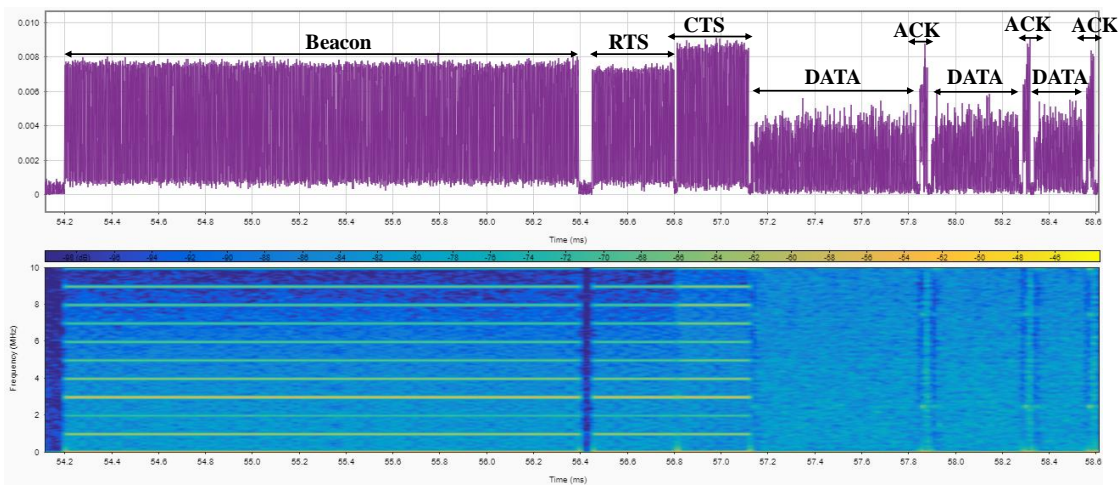

**Fig.5** Wi-Fi signal for data transfer status in the time and frequency-time domain.

### C. Sampling Bluetooth signal

For 802.15 signal sampling, we need at least two devices: a transmitter (MASTER) and a receiver (SLAVE). The transmitter generates a Bluetooth signal and the receiver connects to it so that it can receive the data that the transmitter sends. Bluetooth hops to 79 channels by performing the frequency-hopping spread spectrum (FHSS) technique. These frames can be sent depending on the length of the data by occupying 1, 3, or 5 time slots. Figure 6(a) shows the sampled signal in GNU Radio software. The Figure 6(b) shows Bluetooth frames in MATLAB in the time domain. The sizes of the data frames all occupy 5 time slots, and they are between 2500 and 2870$\mu$s, according to Table 1. The receiver transmits the ACK frame to the transmitter after receiving the data over a period of 200 to 600$\mu$s to inform the transmitter that the data frame was successfully received. The time frame of the ACK frame, as shown in Table 1, because it occupies a one-time slot, varies between 126 and 360$\mu$s. It should be noted that we set the sampling rate at 20MS/s, and therefore, we can monitor only about 1/4 of the ISM band. But remember that our target is to detect and classify the Bluetooth signal (and not to extract data), and this can be achieved by monitoring some parts of the signal in the frequency domain.

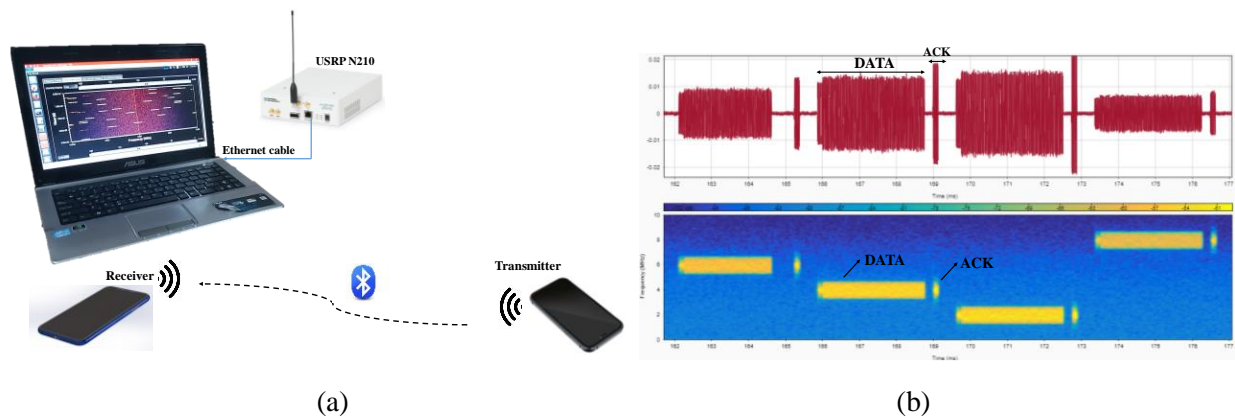

(a)          (b)

**Fig6**. (a) The experiment for sampling Bluetooth. (b) The Bluetooth signal in MATLAB in time and time-frequency domains.

Finally, in another scenario, we consider the status in which both Wi-Fi and Bluetooth signals are presented on the air at the same time while the Wi-Fi is in the beacon sending status. The system setup and time and time-frequency presentation of a part of the signal are shown in Figure 7.

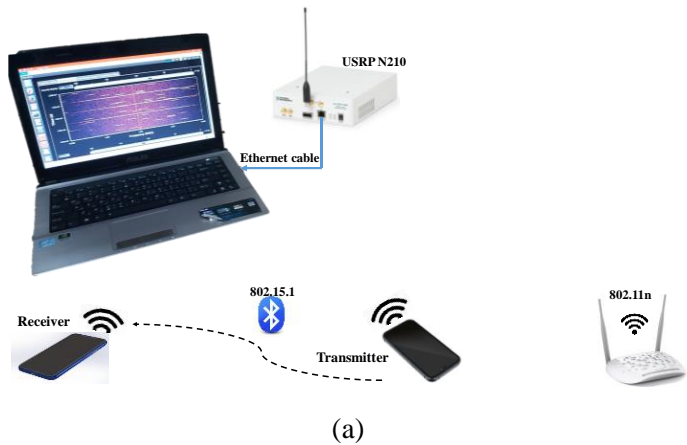

(a)

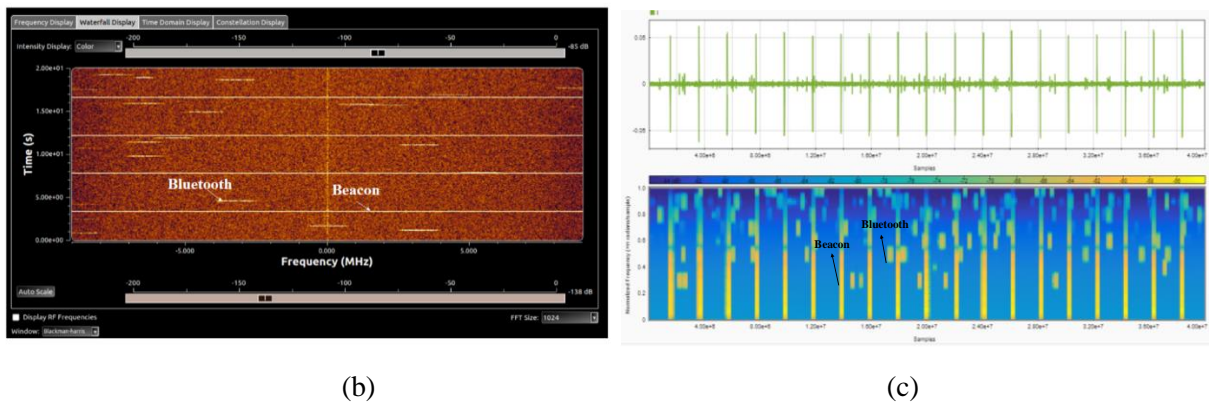

(b)             (c)

**Fig7**. (a) The experiment for sampling Wi-Fi Beacon and Bluetooth. (b) Both Wi-Fi Beacon and Bluetooth signals are on GNU Radio. (c) Both signals in MATLAB in the time and time-frequency domains.

## IV. Feature Extraction and Classification

In this section, we do the basics of classifying protocols. For this purpose, after preparing the samples, we extract the features. We must manually extract the features to use the SVM and KNN methods to classify protocols. We extract the frame width and the silence gap between the two frames as time features and the PAPR of each frame as a power feature.

### A. Time Feature extraction

Time, frequency, power, or a combination of these domains can be used to identify and classify protocols, similar to [4] and [11]. Reference [4] utilizes both time and power features to classify simulated protocols, while reference [11] only uses time features. By analyzing the sampled signals in the time domain, we observe that we can obtain useful information from the protocols. By detecting frames, we can get the frame width and the silence gap between the two frames. These two features are shown in Figure 8 for a sample signal to help us comprehend them better.

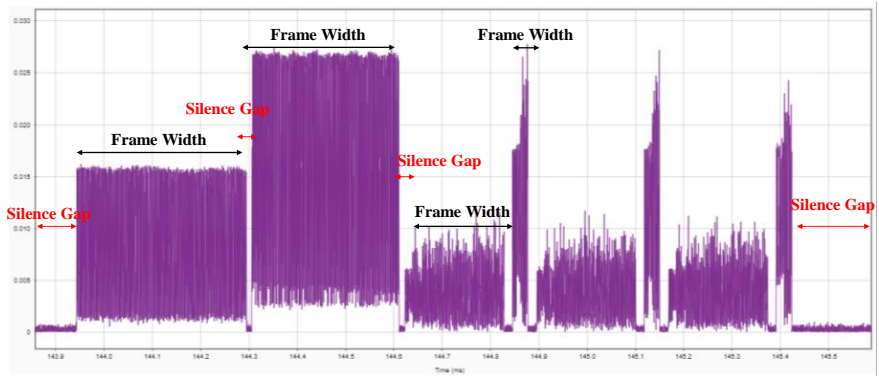

(a)

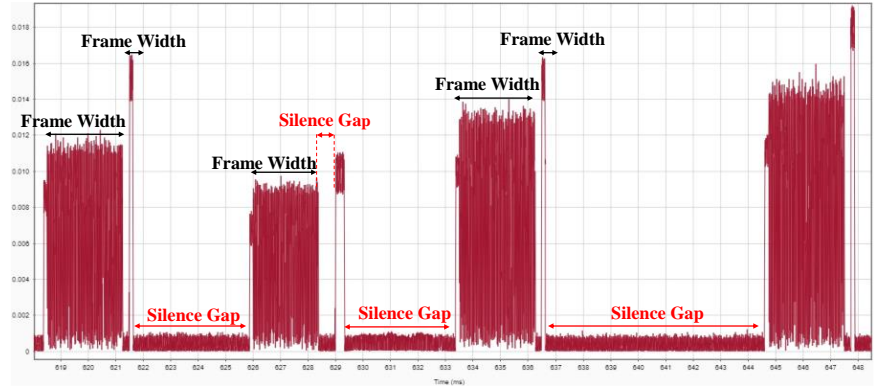

(b)

**Fig8**. (a) Features of Wi-Fi signal in the time domain. (b) Features of Bluetooth signal in the time domain.

The frame width is obviously the duration from the rising edge (the start of the frame) to the falling edge (the end of the frame). In the 802.11n signal, the beacon, RTS, CTS, and ACK frames, and in the Bluetooth signal, the data and ACK frames all have almost constant values. The duration of silence gaps between two successive frames can be considered as the second feature. This value is the difference between the frame's start time and the previous frame's end time. Recalling Figure 1, the SIFS and DIFS durations, which are 10 and $50 \mu s$, respectively, are the silence gaps between the RTS, CTS, Data, and ACK frames in the 802.11n signal.

We must first detect the frames to extract these two features. For this purpose, the rising and falling edges of the frame must be detected. There are several methods for detecting frames in the signals. We consider the technique of blind detection of burst signals, as presented in [23], and propose a modified version of it. Using the above method, we calculate the signal power and pass it through a softener filter after receiving the input signal. The weight of the left and right halves of the window is then calculated using the slider window and Equation (14), and the rising and falling edges are detected. The energy ratio function at the $K^{\text{th}}$ point of the signal can be represented as follows to detect the rising and falling edges of the frame:

$$E_W(K) = \frac{\sum_{i=K-L+1}^{K}|r_i|}{\sum_{i=K+\Delta}^{K+\Delta+L-1}|r_i|} \begin{matrix} < \\ > \end{matrix} \alpha \qquad (14)$$

Our modification is adding a gap between the left and right half-window of size Δ samples. Windows move over the signal. Figure 9 shows the window at point *K*. The value of α defines the threshold level, which is a constant number (we set $\alpha = 2.7$). The width of the windows is denoted by the letter *L* (the size of the windows for the rising and falling edges may be different). Based on the occurrence of rising and falling edges, these edges are determined by comparing $E_W(K)$ with the threshold level α. After finding the falling and rising edges in the signal, we can easily find the frames and the gaps between consecutive frames.

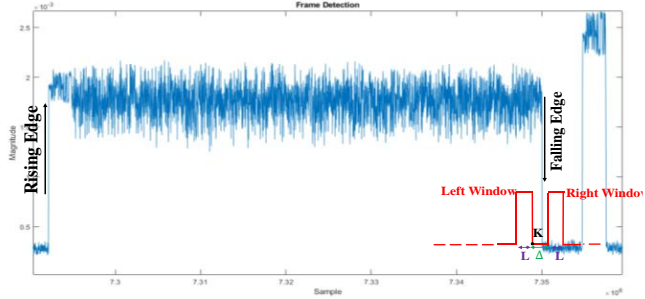

**Fig.9** The sliding window movement to detect rising and falling edges after the input signal passes through the softener filter.

## B. Power Feature Extraction

In the OFDM signal, when the subcarriers are modulated independently, they are in different phase values relative to each other at any given moment. As a result, when these subcarriers achieve their maximum phase value at the same time, suddenly a big peak occurs. Thus, the peak value of the signal compared to the average of the total signal can be quite high. This causes an OFDM system's peak power consumption to be extremely high, which is one of the OFDM disadvantages [24]. As a result, this property can be used as a feature to identify the OFDM signal from other signals. Peak to Average Power Ratio (PAPR) of a signal *x(t)*, is defined as,

$$\boldsymbol{PAPR} = \frac{max[|x(t)|^2]}{E[|x(t)|^2]}, t \in [0, T_s] \tag{15}$$

802.11n employs OFDM signaling in the physical layer, and therefore, PAPR is relatively more in this protocol, compared to Bluetooth. As a result, we consider the PAPR of the detected frames as the power feature, which helps us to classify the protocols more accurately.

Figure 10 illustrates the architecture of our design. The system block diagram includes all the steps implemented in this work. We will describe each step in the remainder of this section.

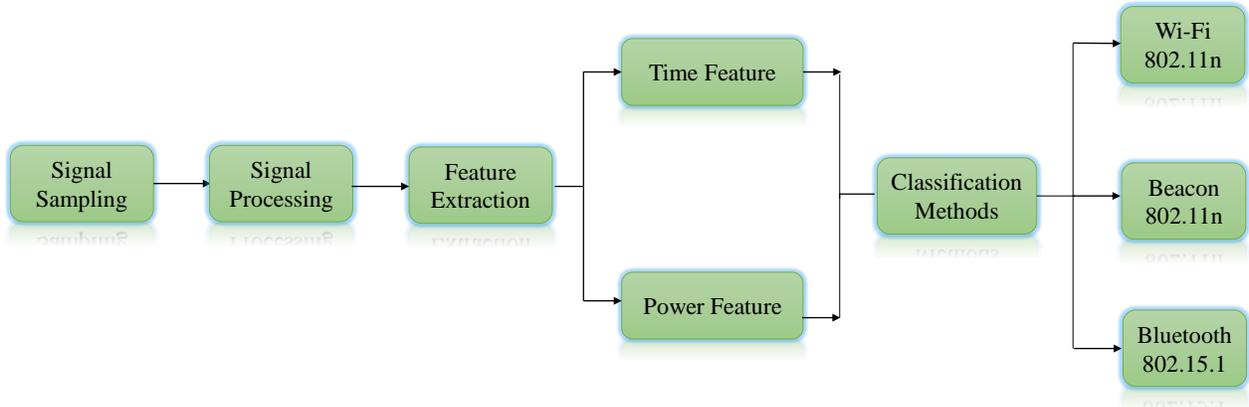

**Fig.10** System block diagram.

## C. Classification by SVM & KNN

First, we use time features to create the data's feature space. We sample the signals in a few seconds, which is approximately 1000 frames per class. We use nonlinear methods to classify the protocols because, as shown in Figure 11, the data appeared to be linearly inseparable in the feature space.

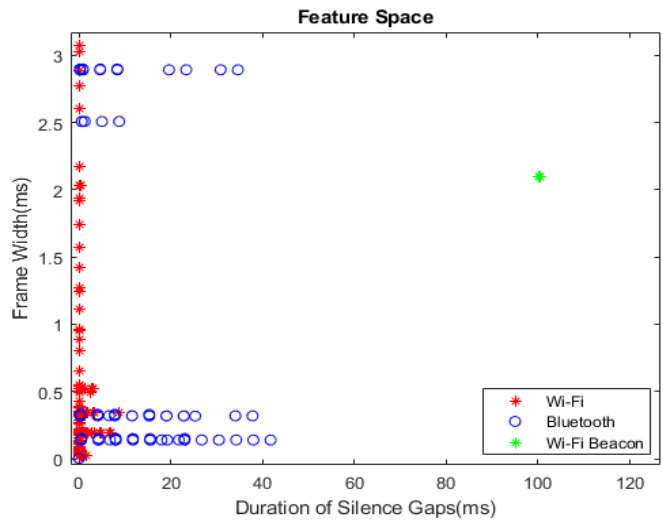

**Fig11.** Bluetooth, Wi-Fi, and Wi-Fi Beacon protocols in feature space

In the classification with the proposed methods, we considered 20% of the data for testing and 80% of the data for training. We use the one-vs-all approach to classify the data using SVM and KNN into three classes: Wi-Fi, Bluetooth, and Wi-Fi Beacon. For example, Figure 12 illustrates the data classification with the Gaussian kernel.

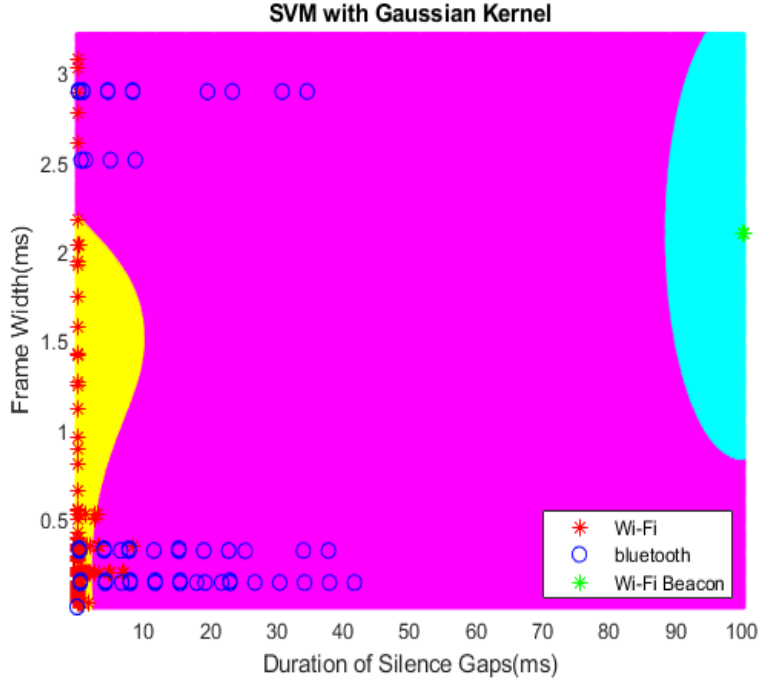

**Figure 12.** Protocol classification by SVM method with Gaussian kernel

As you can see, the data is classified into distinct colors in its region using the Gaussian kernel SVM method, which has an accuracy of 92.15 percent. With linear and polynomial kernels, this value is equal to 78.03 and 88.08, which shows that the Gaussian kernel has the best accuracy.

For KNN classification, we considered the number of neighborhoods to be 10 ($K = 10$). Figure 13 illustrates the neighborhood with a circle around the new test, in which ten neighbors have been determined by Euclidean distances. The classification accuracy with this method is 93.11 percent.

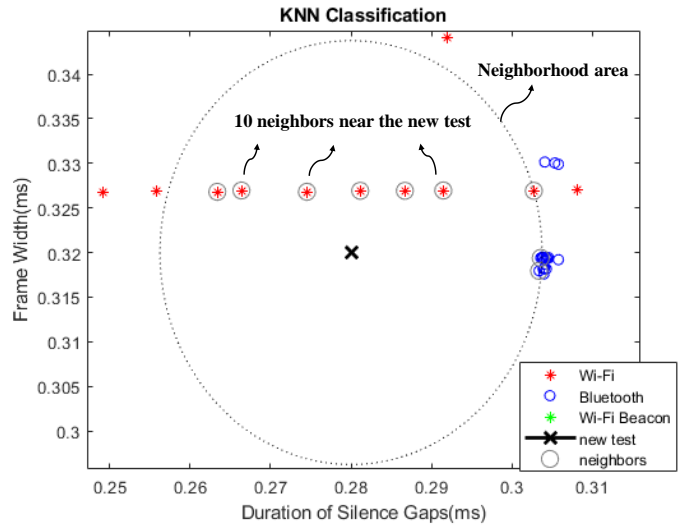

**Figure 13.** 10 Neighbors near the new test in the KNN classification

We add the PAPR power feature for better classification accuracy. As a result, our feature space is transformed into a three-dimensional space. Figure 14 shows that the PAPR of the Wi-Fi signal is higher than the Bluetooth signal as forecasted. Also, in this case, we use SVM and KNN methods to classify the frames into three classes based on three features. Table 4 illustrates the classification accuracy for each method, which has been improved significantly due to the addition of a power feature.

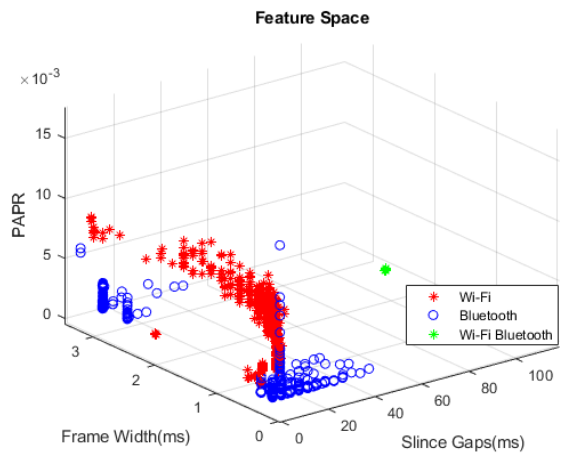

**Figure 14.** Bluetooth, Wi-Fi and Wi-Fi Beacon frames in 3-dimensional feature space

**Table 4.** Accuracy of Classification

| Method of Classification | Classes | Features | Accuracy of Classification |
|---|---|---|---|
| Linear-SVM | Wi-Fi, Bluetooth, Wi-Fi Beacon | Frame Width, Silence Gaps, PAPR | 95.15% |
| Polynomial-SVM | Wi-Fi, Bluetooth, Wi-Fi Beacon | Frame Width, Silence Gaps, PAPR | 96.21% |
| Gaussian-SVM | Wi-Fi, Bluetooth, Wi-Fi Beacon | Frame Width, Silence Gaps, PAPR | 97.83% |
| KNN | Wi-Fi, Bluetooth, Wi-Fi Beacon | Frame Width, Silence Gaps, PAPR | 98.12% |

**D. Classification of protocols in noisy conditions**

In this part, we look at how well feature extraction and classification methods perform in the presence of additive white Gaussian noise (AWGN). The training data are the same as the previous signals, while the test data are signals that are manually set to variable SNRs. The sampled signal has a small amount of Gaussian noise and is also affected by the multi-path event. We investigate the effect of increasing Gaussian noise via simulation, as it is too hard to gather the signal at varied distances and situations with different SNRs. Our guess is that the effect of the multi-path event on our work is not significant, as it does not cause noticeable changes in the frame width and silence gaps. This issue and its effects can be investigated in detail in future works.

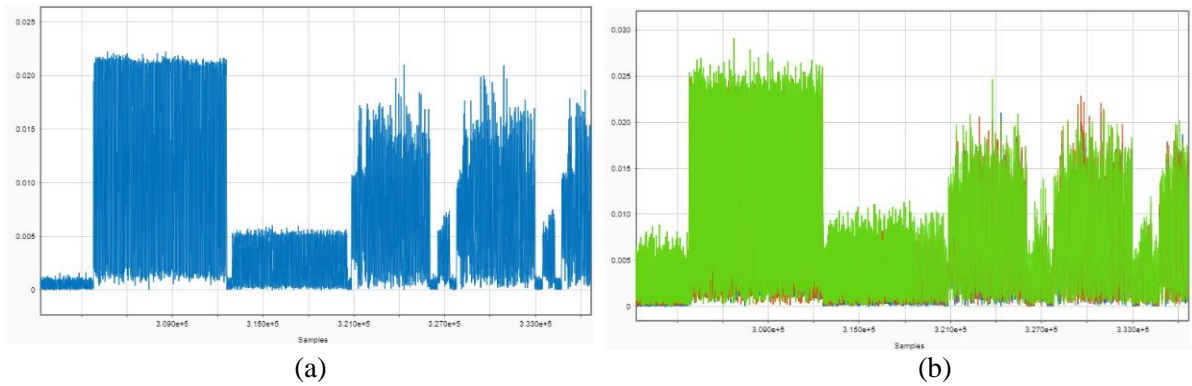

(a) (b)

**Figure 15.** (a) (Almost) Noise-free Wi-Fi signal. (b) Wi-Fi signal with SNR = 5.

Figure 15 illustrates the exterior difference between the noisy signal and the noise-free signal. When we compare two pictures in Figure 15, we can see that frames with lower amplitudes vanish as the noise power levels rise. Thus, detecting weaker frames becomes more difficult. As a result, the classification accuracy decreases due to the reduced number of detected frames or even false frame detection alarms.

As we said, the first step of our classification process is frame (or burst) detection. For the actual signal, which we assume to be noise-free with some tolerance, the number of detected frames using the presented algorithm is known. The performance of the frame detection step is dropped, as depicted in Figure 16.

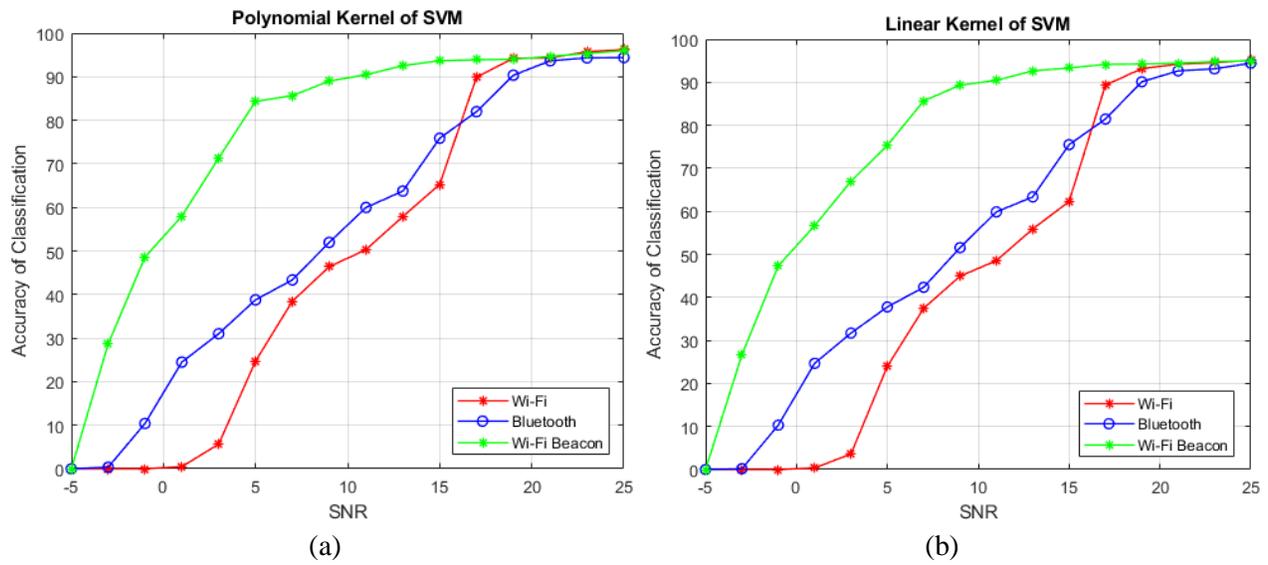

(a) (b)

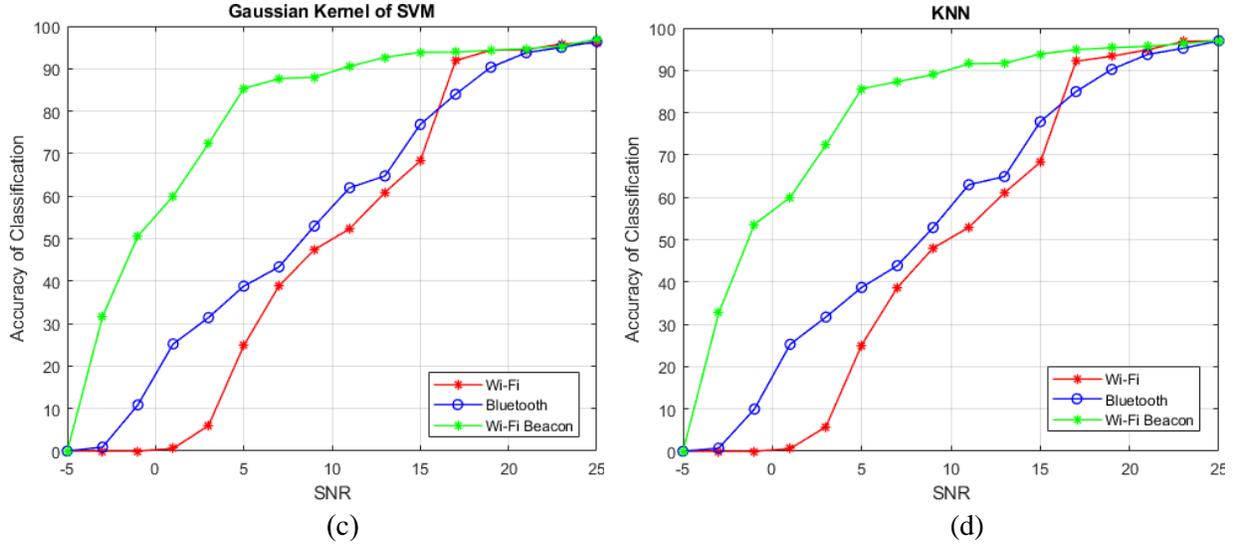

**Fig16.** (a) Classification accuracy in noisy conditions using SVM method with linear kernel, (b) SVM method with polynomial kernel, (c) SVM method with Gaussian kernel, (d) KNN method.

After that, we detected the frames in the signal. We want to see if the frames detected in a test signal belong to the same class or not, so we check in several methods. For example, we consider a Wi-Fi signal with 850 frames as one of the test signals for the system. In this case, we set SNR = 5. Now, the number of frames detected in this test signal is calculated to be 217. We examine all four methods and calculate the classification accuracy to see how many of the 217 detected frames are correctly classified in the Wi-Fi class. Based on the accuracy of the classifications in Table 4 in normal, each method classifies some of these 217 frames into the Bluetooth and Beacon classes. The SVM method with a linear kernel successfully classifies 212 frames in the Wi-Fi class and incorrectly classifies the remaining 5 frames in other classes. A total of 217 frames are accurately classified in the Wi-Fi class using the Gaussian kernel and KNN methods. By comparing the diagrams in Fig. 16, as expected, the two methods of Gaussian kernel and KNN had the highest classification accuracy, and the accuracy of the methods has grown as the SNR has increased.

## V. Conclusions and future work

This paper embarked on a mission to classify MAC sublayer protocols, specifically focusing on Wi-Fi and Bluetooth, using machine learning methods. Driven by the burgeoning advancement in machine learning applications within wireless networks and cognitive radio systems, our study incorporated these sophisticated techniques to effectively discern and classify the said protocols.

Through a rigorous methodology of signal sampling and feature extraction, focusing on time and power domains, we optimized Support Vector Machines (SVM) and K-nearest neighbors (KNN) classifiers. Our findings elucidate that the SVM with a Gaussian kernel and the KNN classifier eclipsed in performance, yielding remarkable classification accuracies of 97.83% and 98.12%, respectively. These classifiers exhibited profound resilience, adeptly navigating through the challenges posed by additive white Gaussian noise (AWGN), underscoring their robustness and reliability.

Peering into the future, this research heralds a pathway for exploring diverse wireless network protocols beyond those examined in this study. The horizon beams with potential, inviting exploration of advanced methodologies such as deep learning algorithms and neural networks for more nuanced and robust protocol classification strategies.